\documentstyle[12pt]{article}

\begin{document}

{\sc
\centerline{ Finite $3\pi$ cut approximation for the
                $\pi N\bar{N}$ form factor}
}
\vskip1.cm
\centerline{ A.A.Bolokhov }
\centerline{\sl Sankt-Petersburg State University, }
\centerline{\sl Sankt-Petersburg, 198904, Russia }
\centerline{ and }
\centerline{ N.Zovko }
\centerline{\sl Ruder Bo\v{s}kovi\'{c} Institute, }
\centerline{\sl Zagreb, Croatia }

\vskip2.cm
\centerline{\bf Abstract }
\vskip1.cm

                Assuming the length of the
        $ 3\pi $
                cut
                to be finite and approximating the integrated
                amplitude by a constant,
                we derive an expression for the
        $\pi N\bar{N}$
                form factor which is very close to that
                given by a simple pole.
                The specific predictions of the obtained form
                factor for the region of small momentum
                transfer are discussed along the lines of the
                Goldberger--Treiman relation.


\vskip2.cm
1. INTRODUCTION
\vskip1.cm

                The problem of determining the
        $\pi N\bar{N}$
                vertex with the pion off its
                mass shell has been
                open for about 30 years and as yet has
                had no reliable solution. There exist a few
                model calculations for zero-mass pions or very
                close to this limit. Being designed for the
                space--like region only,
                these models have little
                chance for any extrapolation.

                Some new trends in the pion--nucleon physics
                were triggered by the claim of the Nijmegen group
                (de Swart and coworkers) for a lower value of the
                pion--nucleon coupling constant
\cite{Nij}
                .
                This claim was later supported by the VPI analysis of
                Arndt and coworkers
\cite{VPI}
                , and
                might initiate a large scale revision of all
                data on pion--nucleon
                phenomenology since the absolute value
                of the pion--nucleon coupling constant
        $g_{\pi N}$
                enters almost all
                analyses (mostly in the fixed input
                form).

                Whereas the method of Ref.
\cite{Nij}
                for extracting the coupling constant based on
                the asymptotic behaviour of the one--pion
                exchange potential in configuration
                space was declared to be insensitive to the
                shape of the form factor (FF), the very
                properties of the
        $\pi N\bar{N}$
                vertex had to be
                taken into account
                in a general determination of the
        ${\pi N}$
                coupling constant. This was confirmed in Ref.
\cite{soft}
% {soft} J.Haidenbaucr, K.Holinde and A.W.Thomas. Phys.Rev.
% {\bf C45}, 952(1992).
                by comparison of two sets of
                one--boson--exchange models with the conclusion
                that the choice of either a softer
        ${\pi N}$
                form factor  or a smaller
        ${\pi N}$
                coupling constant improved the fit.

                On the other hand, the criticism of the Nijmegen
                result by Ericson
\cite{Er}
                is focused on the specific choice of an
                exponential FF made in Ref.
\cite{Nij}
                . The conclusion of Ericson simply means that
                if nature favours the monopole-type FF,
                then
                the authors of Ref.
\cite{Nij}
                must have found that their results should be
                sensitive to the choice
                of the form-factor parameter.
%                Basing on the finite cut approximation
                We shall
                see later on that, in fact, the shape of the
        $\pi N\bar{N}$
                form factor should be very close to that of the
                monopole type.

                One will find a similar degree of sensitivity
                to the FF shape in any problem involving the
        ${\pi N}$
                coupling constant.
                Therefore, any attempt to clarify the structure
                of the
        ${\pi N}$
                form factor is being timely and of importance.

                In what follows we present the derivation of
                a modified pion--nucleon FF based on its
                analyticity
                properties in the time--like region where the
%        $\pi N\bar{N}$
                form factor
                develops an imaginary part (Sec. 2).
                While performing the
                integration over the three--pion cut, we assume
                that the off-shell behaviour of
        $4\pi$
                and
        $3\pi N\bar{N}$
                vertices
                is determined by the existence of an
                effective cut of finite length.
                Then approximating the
                contribution of the
        $4\pi$ vertex
                simply by a constant, we arrive at the
                expression for the form factor
                (Secs. 3, 4).
                Its specific features
                are mostly of kinematical origin.
                They are discussed in Secs. 5, 6.

\vskip2.cm
2. THE TIME--LIKE REGION
\vskip1.cm

                 In the time--like region there is no
                experimental information about the pion--nucleon
                form factor.
                Therefore the main formula of
                dispersion theory,
\begin{equation}
        G(\tau) = \frac{1}{\pi} \int_{\tau_{0}}^{\infty} d\tau'
                        \frac{ Im \, G(\tau')}{\tau' - \tau}
                        \; ,
\end{equation}
                (as well as any variant with the suitable
                number of subtractions)
                was practically of no use.

                Already the first unitarity diagram for the
                imaginary part contains an unsurmountable
                difficulty in handling the inelastic
                amplitudes. However, the situation might seem
                less desperate if we observe that the
        $N\bar{N}$
                pair may be considered as an effective pion
                very far away from its mass shell
        $(m_{\pi^*}\approx 2\; GeV)$
                .
                Intermediate pions are physical
                pions by the very fact that in unitarity
                diagrams the intermediate set of states is
                always on the mass shell. So, in essence, the
                evaluation of the imaginary part will include
                the product of two pion-pion elastic scattering
                amplitudes, each of them containing one pion off
                the mass shell.

                Let us now attack the
        $3\pi$
                cut along these lines.
                The
                conventional expression for the
        $\pi N\bar{N}$ vertex
                in the
        $N\bar{N}$
                annihilation region has the form
\begin{eqnarray}
        & & <N_{\alpha}^{(\mu)}(p)\bar{N}_{\beta}^{(\nu)}(p')
                |S-1|\pi^{d}(q)>    \nonumber\\
        & & =i(2\pi)\delta^{(4)}(p+p'-q)
                \times \bar{v}^{(\nu)}(p')i\gamma_5
                u^{(\mu)}(p)\tau_{\alpha\beta}^d \,
                G_{\pi N\bar{N}}(\tau),
\end{eqnarray}
                where
\begin{equation}
        G_{\pi N\bar{N}}(\tau) = g_{\pi N} \cdot G(\tau) ,
        \; \tau \equiv q^2  \, ,
\end{equation}
                with
        $g_{\pi N}$
                being the pion--nucleon coupling constant and
                the form factor
        $G(\tau)$
                is normalized by
\begin{equation}
        G (m_{\pi}^2) = 1 ;
\end{equation}
                other notation is obvious.

                The unitarity condition for
        $S=1+iT$
                , after straightforward
                manipulations, in the
        $3\pi$--cut
                approximation has the form
\begin{eqnarray} & &
2\pi\delta^{(4)}(p+p'-q)\bar{v}^{(\nu)}(p')i\gamma_5
u^{(\mu)}(p)\tau_{\alpha\beta}^d\,[G_{\pi
N\bar{N}}(\tau)-G_{\pi N\bar{N}}^{*}(\tau)]
                                        \nonumber \\
& & =\sum<N\bar{N}|T|n><n|T^{\dagger}|\pi^d>
                                        \nonumber \\
& & \approx\int\prod\frac{d^3k_i}{(2\pi)^{3}2k_{0i}}
<N_{\alpha}^{(\mu)}(p)\bar{N}_{\beta}^{(\nu)}(p')|T|
\pi_a(k_1)\pi_b(k_2)\pi_c(k_3)>
                                        \nonumber \\
& & \times<\pi_a(k_1)\pi_b(k_2)\pi_c(k_3)|T^{\dagger}|
\pi^d(q)>
                                        \nonumber \\
& & =\int\prod\frac{d^3k_i}{(2\pi)^{3}2k_{0i}}
(2\pi)^4\delta^{(4)}(p+p'-k_1-k_2-k_3)
M_{abc}^{\alpha\beta,\mu\nu}(N\bar{N}\rightarrow 3\pi)
                                        \nonumber \\
& & \times(2\pi)^4\delta^{(4)}(q-k_1-k_2-k_3)M_{abcd}^{*}
(\pi\rightarrow 3\pi)
                                        \nonumber \\
& & =(2\pi)^4\delta^{(4)}(p+p'-q)\int d\Omega_{3\pi}
M_{abc}^{\alpha\beta,\mu\nu}(N\bar{N}\rightarrow 3\pi)
M_{abcd}^{*}(\pi\rightarrow 3\pi),
\end{eqnarray}
                where
\begin{equation}
\int \prod \frac{d^3k_i}{(2\pi)^32k_{0i}} \times
        \delta^4(q - k_1 - k_2 - k_3)
\equiv \int d\Omega_{3\pi}.
\end{equation}
                After dropping the
        $\delta$--functions,
                the unitarity condition in the
        $3\pi$
                -intermediate-state approximation becomes
\begin{equation}
2\bar{v}^{(\nu)}(p')i\gamma_5u^{(\mu)}(p)
\tau_d^{\alpha\beta} Im \, G_{\pi N\bar{N}}(\tau)
= \int d\Omega_{3\pi}M_{abc}^{\alpha\beta,\mu\nu}(N\bar{N}
\rightarrow 3\pi)M_{abcd}^{*}(\pi\rightarrow 3\pi).
\end{equation}
                In order to form an isospin pion state in the
                left-hand side, we have to multiply it by the
                factor
        $\tau_{d'}^{\beta\alpha}
                $
                .
                The antiparallel-spin nucleon wave functions
                should be used for representing the pion state.
                For this purpose, we simply multiply  both sides by
\begin{equation}
        \bar{u}^{(\mu)}(p)
        i\gamma_5
        v^{(\nu)}(p')
\end{equation}
                and perform summation over
        $\mu , \nu$
                to remove the spinor structures.
                Denoting
\begin{equation}
                M_{abcd'} \equiv
                \frac{1}{\tau}
                \frac{1}{2}
                \sum_{\beta , \alpha}
                \frac{1}{4}
                \sum_{\mu , \nu}
                M_{abcd}^{\alpha \beta , \mu \nu}
                 (N\bar{N} \rightarrow 3\pi)
                \times
        \bar{u}^{(\mu)}(p)
        i\gamma_5
        v^{(\nu)}(p')
                \tau_{d'}^{\beta\alpha},
\end{equation}
                we finally obtain
\begin{equation}
        Im \, G_{\pi N\bar{N}}(\tau)\delta_{dd'} =
        \frac {1}{2}
        \int d\Omega_{3\pi} M_{abcd'} M_{abcd}^{*}.
\end{equation}

                Here we should note that the most important
                property of our FF comes from the
        $(1/\tau)$
                multiplier in the expression for
        $Im \, G_{\pi N\bar{N}}$
                .
                One should stress
                that it is not the
                factor
        $(1/\tau)$
                in expression (9).
                Indeed, the
                latter must be contracted in the major spinor
                structure of the vertex
        $M_{abcd}^{\alpha \beta , \mu \nu}
            (N\bar{N} \rightarrow 3\pi)$
                in the explicit form, and in all the other
                structures after the
        $d\Omega_{3\pi}$
                integration.

\vskip2.cm
3. INTEGRATION OVER $3 \pi$ PHASE SPACE
\vskip1.cm

                Let us now consider the integration over the
        $3 \pi$
                phase space.  Following the standard
                definitions and conventions, we first convert
                the integration over the internal pion momenta
                into the integration over the scalar invariant
                variables
\cite{Byck}:
\begin{equation}
        \int d\Omega_{3\pi}
                 = \frac{1}{(2\pi)^5} \frac{\pi}
                                {2^4\lambda^{\frac{1}{2}}
                (\tau,m_N^2,m_N^2)}\int
                \frac{ds_1ds_2dt_1dt_2}
                     {\sqrt{-\Delta_4}}
                     ,
\end{equation}
                where
        $\Delta_4$
                is the Gram determinant \\
\begin{equation}
\Delta_4\equiv\Delta_4(p,p',k_1,k_3)=
\left|\begin{array}{clcr}
p\cdot p   &  p\cdot p'  &  p\cdot k_1   &  p\cdot k_3  \\
p'\cdot p  &  p'\cdot p' &  p'\cdot k_1  &  p'\cdot k_3  \\
k_1\cdot p &  k_1\cdot p'&  k_1\cdot k_1 &  k_1\cdot k_3 \\
k_3\cdot p &  k_3\cdot p'&  k_3\cdot k_1 &  k_3\cdot k_3
\end{array}  \right|
\end{equation}
                and the scalar variables are those used in
                Ref.
\cite{Byck}
\begin{eqnarray}
        s_1 = (k_1 + k_2)^2 \, ,
        s_2 = (k_2 + k_3)^2 \, ,
        \tau = (p + p')^2 \, , \nonumber \\
        t_1 = (p - k_1)^2 \, ,
        t_2 = (p' - k_3)^2 \, .
\end{eqnarray}
                The region of integration is limited to that
                where
        $\Delta_4\leq 0$
                and
        $q^2\equiv \tau \geq 9m_{\pi}^2$.

                We shall take advantage of introducing
                two other variables and renaming the rest.
                In terms of the relative nucleon momentum
        $P = p - p' $,
                the suitable invariant variables are
\begin{equation}
        t'_1 = P\cdot (k_1+k_3) , \;\;
        t'_2 = P\cdot (k_1-k_3) , \;\;
        t \equiv s_1 , \;\; s \equiv s_2 , \;\; \tau .
\end{equation}
                Here $s$ and $t$ are the usual Mandelstam
                variables of the
        $4\pi$ vertex
                and $\tau$ is the mass of the heavy pion.

                There are two important properties of the
        $t'_1$ , $t'_2$
                variables. First, the
        $4\pi$ vertex
                does not depend on them. There are also grounds
                to consider the dependence of the amplitude
        $M (N\bar{N} \rightarrow 3\pi)$
                on these variables to be negligible: the full
                kinematics data
\cite{BLO63}
% {BLO63} T.D.Blokhintseva et al. JETP, 44 (1963) 498.
\cite{JON74}
% {JON74} J.A.Jones, W.W.Allison and D.H.Saxon.
% Nucl. Phys. B83 (1974) 93.
                on the cross reaction
        $\pi^{+} p \rightarrow \pi^{+} \pi^{-} n$
                show
                no dependence on these variables
                apart from
                the dependence coming from the phase space.
                The analysis
\cite{Lowe}
                of recent measurements also confirms this
                observation.

                Thus we can rely upon the fact that the
                integrated expression in (11) is free of
                explicit
                $t'_1$ , $t'_2$
                dependence.

                Second, the integration domain of
                $t'_1$ , $t'_2$
                is an ellipse in the
                $t'_1$ , $t'_2$
                plane,
                so the above integration can be easily performed.

                Finally, our $3\pi$-phase-space
                integral (11) takes the form
\begin{equation}
        \int d\Omega_{3\pi}=
        \frac{const}{\tau}\int_{S_0}^{S_1}ds\int_{t^-}^{t^+}dt,
\end{equation}
where
\begin{equation}
 S_0=4m_{\pi}^2,
\end{equation}
\begin{equation}
        t^{\pm}=\frac{1}{2}
                \left[ \tau+3m_{\pi}^2-s \pm
                \frac{\sqrt{(s-4m_{\pi}^2)(s-S_1)(s-S_2)}}
                                {\sqrt{s}}
                \right],
\end{equation}
\begin{equation}
S_{1,2}=(-\sqrt{\tau}\pm m_{\pi})^2.
\end{equation}

                This result contains all
                $q^2\equiv \tau$
                dependence in the approximation when the
                amplitude $M$ in the integral (10)
                is taken to be a constant.

\vskip2.cm
4. BEYOND THE EFFECTIVE POLE APPROXIMATION
\vskip1.cm

                If the effective pole formula
                (see eq. (23) below )
                tolerates
        $\lambda^2 \approx \tau_0
                \equiv 9m_{\pi}^2$
                , it means that the form-factor behavior is
                determined by the very fact of the presence of
                the cut from
        $\tau_0$
                to infinity.
                If the structure of the amplitude
        $M(\pi_{*}\rightarrow 3\pi)$
                in the eq.(10) is essential too, then
        $\lambda^2 \gg \tau_0$
                .

                 In this case it might be convenient to
                introduce the notion of the effective cut-off
         $\tau_m$
                in the amplitude
         $M(\pi_{*}\rightarrow 3\pi)$
                such that any integration of the amplitude
                squared is equivalent to the finite-range
                integration of some mean value.
                If this holds, then a meaningful approximation
                could be obtained in the following way:
\begin{eqnarray}
        G_{\pi N\bar{N}}(\tau) &=& K \int_{\tau_0}^{\infty}
                \frac{d\tau '}{\tau '-\tau}
                \frac{1}{\tau '} \int\int_{D(\tau')} ds dt
        \, |M(\pi_{*}\rightarrow 3\pi)|^2  \nonumber  \\
        & \approx & K\int_{\tau_0}^{\tau_m}
                \frac{d\tau '}{\tau '-\tau}
                \frac{1}{\tau '} K'
        = \frac{KK'}{\tau}
        \ln\frac{1 - {\tau}/{\tau_m}}
                 {1 - {\tau}/{\tau_0}} ,
\end{eqnarray}
                where the overall constant $KK'$ must be
                determined by the normalization,
        $G_{\pi N\bar{N}}(m_{\pi}^2) = g_{\pi N}$.

                If the double integral including the factor
        $ 1 / \tau' $
                is approximated by the constant,
                we obtain the (normalized)
                form factor
\begin{equation}
         G(\tau) = \frac {\ln [
                 ( \tau_m - \tau ) /
                 ( \tau_0 - \tau ) ] }
                         {\ln [
                 ( \tau_m - m^2_{\pi} ) /
                 ( \tau_0 - m^2_{\pi} ) ] }
                 .
\end{equation}
                If the value of the cut-off parameter
        $ \tau_m $
                is close to
        $ \tau_0 $
                ,
                then the FF (20) and the FF we shall be dealing
                with later on are practically the same and are
                very close to the
                form of a simple pole.

                In the limit
        $ \tau_m \rightarrow \infty $,
                expression (20) is a constant
        $ ( = 1 ) $
                .
                When approximating the inner integral in (19)
                by a sequence of
        $\Theta$--function
                times some simpler expansion,
                the weight of the term giving
                rise to expression (20) must vanish
                at large cut-off. Otherwise it will give rise
                to the form factor containing a hard core.
                In our oppinion, one would like to exclude this
                possibility.

                The approximation we have made is very rough, but
                it leaves us with a single free parameter
        $\tau_m$
                .
                One can realize that inclusion of
                further dynamical details  immediately
                converts the form factor problem into a
                multiparametric one.

                Therefore, in the approximation of a constant
                amplitude over the cut of finite length,
                our normalized form factor assumes the form
\begin{equation}
         G(\tau) = \frac{g_0}{\tau} \ln \frac
                 {1 - {\tau} / {\tau_m}}
                 {1 - {\tau} / {\tau_0}} ,
%                 {1-\frac{\tau}{\tau_m}}
%                 {1-\frac{\tau}{\tau_0}} ,
\end{equation}
                with
\begin{equation}
         g_0 = \frac{m^2_{\pi}} {\ln \frac
                 {1 - {m^2_{\pi}} / {\tau_m}}
                 {1 - {m^2_{\pi}} / {\tau_0}} } .
\end{equation}

                  The most important properties of the expression
                  (21), to be pointed out immediately, are

                1)
        $ \tau = 0 $
                is not a singularity;

                2) for
        $\tau \rightarrow -\infty$,
                it decreases as
        $ 1 / \tau $
                ,
                 as in the case of a simple pole.

                In what follows we investigate the
                properties of the FF (21) mainly by comparing
                it with those of the well-known
                form of a (simple) pole,
\begin{equation}
         P(\tau) =  \frac
                 {\lambda^2 - m^2_{\pi}}
                 {\lambda^2 - \tau}  .
\end{equation}
                We leave aside the detailed comparison with
                other broadly used kinds of FF's such sa a dipole,
\begin{equation}
         D(\tau) =  \frac
                 {( \Lambda^2 - m^2_{\pi} )^2}
                 {( \Lambda^2 - \tau )^2}
\end{equation}
                or an
                exponential,
\begin{equation}
         E(\tau) =  \exp \left(  \frac
                 { \tau - m^2_{\pi} }
                 { \Lambda^2_e }
                        \right)
                        .
\end{equation}
                It is a simple algebraic
                exercise to identify the parameters of the FF's
                in question at
        $ \tau \approx 0 $ .
                On the other hand, of the quoted FF's, only
                (21) and the monopole are alike in the
                asymptotic region
        $\tau \rightarrow -\infty$
                .

                 The origin of similarity of these two FF's
                 might be clarified by considering the limiting
                 case
        $\tau_m \rightarrow \tau_0 $
                .
                In this limit, because of the normalization
                condition (4), the
                FF (21) assumes the form
\begin{equation}
         \lim_{\tau_m  \rightarrow \tau_0}
                \frac{g_0}{\tau} \ln \frac
                 {1 - {\tau} / {\tau_m}}
                 {1 - {\tau} / {\tau_0}}
                 =  \frac
                 {\tau_0 - m^2_{\pi}}
                 {\tau_0 - \tau}.
\end{equation}
                So the effective
                monopole shape of the pion--nucleon FF
                might indeed be the natural one and its
                effective mass is
\begin{equation}
                \lambda = \sqrt{\tau_0} = 3m_{\pi}
                        \approx 0.414 {\rm \; GeV } .
\end{equation}

                The most important regions for comparison of
                the FF's in question when
         ${\tau_m  \neq \tau_0}$
                are the region of small
        $ \tau $
                and the asymptotic region when
        $\tau \rightarrow -\infty$
                . Let us now proceed along these lines.

\vskip2.cm
5. $\pi N \bar{N}$ FORM FACTOR AT SMALL MOMENTUM TRANSFERS
\vskip1.cm

                  The region of small
        $\tau$
                is of great importance since the experimental
                quantities measured here meet the theoretical
                predictions of chiral dynamics. Therefore, we
                compare the FF's (21) and (23) in terms of the
                Goldberger--Treiman relation (GTR)
\cite{GT}
                .

                   Let us first briefly recall the usual
                derivation of the GTR. Identifying the
        $G_{\pi N\bar{N}}$
                form factor in the matrix element of the
                pion--field source
        $j^{a}$
                ,
\begin{equation}
        <N(p')|j^{a}(0)|N(p)> =
                iG_{\pi N\bar{N}}(q^2)
                \bar{u}(p')\gamma_5\tau^{a}u(p),
\end{equation}
\begin{equation}
         p' = (p + q) , \; \;
                p'^2 = p^2 = m_N^2 , \; \;
                        q^2 = 2 p \cdot q ,
                        \nonumber
\end{equation}
                one converts, in the standard way, the matrix
                element (28) of the pionic source
        $j^{a}(0)$
                into the pion field
                and then into the partially
                conserved axial current as follows:
\begin{eqnarray}
& & <N(p+q)|j^{a}(0)|N(p)> \, = \, (-q^2+ m_{\pi}^2)
                <p+q| \phi^{a}(0) |p>
                                        \nonumber \\
& & =\frac{-q^2+m_{\pi}^2}
        {m^2_{\pi } f_{\pi}\sqrt{2}}
<p+q|\partial j_{5\mu}^{a}|p>
                                        \nonumber \\
& & =\frac{-q^2+m_{\pi}^2}
        {m^2_{\pi } f_{\pi}\sqrt{2}} \bar{u}(p+q)
i\gamma_5\tau^{a} \, [2m_Ng_A(q^2) + q^2h_A(q^2)]u(p),
\end{eqnarray}
                where the notation is straightforward and
                conventional.

                 After comparing (28) with (30), we obtain
\begin{equation}
        g_{\pi N} G(q^2) =
        \frac{-q^2+m_{\pi}^2}{m^2_{\pi } f_{\pi}\sqrt{2}} \;
                [2m_Ng_A(q^2) + q^2h_A(q^2)].
\end{equation}
                This relation holds in a limited region of
        $q^2$
                where the pion-field-to-axial-current
                identity holds.
                Provided that the axial form factors
        $g_A(q^2)$
                and
        $h_A(q^2)$
                can be measured or calculated, it defines
                the off--shell behavior of the pion--nucleon
                coupling. At
        $q^2=0$
                the Goldberger-Treiman relation
                follows,
        $g_{\pi N} f_{\pi} G(0) =
                \sqrt{2} \, m_N m^2_{\pi} g_A(0)$
                .
                We rewrite it for the  intercept
        $G_0 \equiv G(0)$:
                ,
\begin{equation}
        G_0 = \frac {m_N} { F_{\pi}}
                \frac {G_A} { g_{\pi N} } ,
\end{equation}
                where
        $G_A \equiv 2 g_A(0) = 1.261 \pm 0.004$
\cite{Ahr}
                and
        $ F_{\pi} \equiv f_{\pi} \sqrt{2} =
                        ( 92.6 \pm 0.2 ) {\rm MeV} $
\cite{Holst}
                .

                As
        $g_A(q^2)$
                receives no contribution
                from the pion pole, we can evaluate the slope
                of the form factor
        $G(q^2)$ at $q^2=0$
                by using the pion-pole contribution to
        $h_A(q^2)$:
\begin{equation}
        h_A(q^2)|_{\pi\,pole} =
                \frac{f_{\pi}g_{\pi N}\sqrt{2}}
                        {q^2 - m_{\pi}^2},
\end{equation}
                resulting in
\begin{equation}
        G'(0) = \frac{1}{m^2_{\pi}
        f_{\pi}\sqrt{2}} \,
        [2m_Ng_{A}(0) + m_{\pi}^2h_A(0)] +
                \frac{m_N \sqrt{2}}{f_{\pi}} g'_A(0).
\end{equation}
                The GT relation and
        $h_A(0)$
                from (33) make the bracket vanish, so that
                we finally obtain
\begin{equation}
        \frac{G'(0)}{G(0)} = \frac{g'_A(0)}{g_A(0)}.
\end{equation}

                  Expressions (32) and (35) are the tools for
                comparison of FF's at
        $\tau = 0$
                since any physically acceptable FF must provide
                a reasonable value for the discrepancy of the
                GTR
                as well as for the nucleon size. Let us now
                expand the FFs in question around
        $\tau = 0$
                :
\begin{equation}
         P(\tau) =  \frac
                 {\lambda^2 - m^2_{\pi}}
                 {\lambda^2 - \tau} =
                    \frac
                 {\lambda^2 - m^2_{\pi}}
                 {\lambda^2} [ 1 + \tau / \lambda^2 + ... ]
                  ,
\end{equation}
\begin{equation}
         G(\tau) =  g_0
                \frac {\tau_m - \tau_0} {\tau_m  \tau_0}
                 \left[
                 1 + \tau \frac { 1 } { 2 }
                \frac {\tau_m + \tau_0} {\tau_m  \tau_0}
                 + ...
                 \right]
                  ,
\end{equation}
\begin{equation}
         g_0 = \frac{m^2_{\pi}} {\ln \frac
                 {1 - {m^2_{\pi}} / {\tau_m}}
                 {1 - {m^2_{\pi}} / {\tau_0}} }
                 . \nonumber
\end{equation}
                The terms of the zeroth order then give the
                intercepts
        $P_0$
                and
        $G_0$
                of the monopole and the modified FF,
                respectively, entering the GTR:
\begin{equation}
        P_0 =
                \frac {\lambda^2 - m^2_{\pi}} {\lambda^2}
                  ,
\end{equation}
\begin{equation}
         G_0 =
                \frac {\tau_m - \tau_0} {\tau_m  \tau_0}
                \frac{m^2_{\pi}}
                        {\ln \left(
                                \frac
                                 {\tau_0}
                                {\tau_0 - m^2_{\pi}}
                                \frac
                                {\tau_m - m^2_{\pi}}
                                 {\tau_m}
                               \right)
                         }
                 .
\end{equation}

                Here the first difference in the features of
                the considered FF's appears. The monopole FF
                can in principle tune any, even unphysical,
                value of
        $P_0$
                from zero to unity. In contrast to this, expression
                (40) predicts a very narrow interval of
        $G_0$
                values from
        $G_0^0 = 8/9$
                to
        $G_0^{\infty} = (9 \ln (8/9) )^{-1}$:
\begin{equation}
        0.889 \leq G_0 \leq 0.943,
\end{equation}
                spanned under the variation of
        $\tau_m$
                from
        $\tau_0$
                to infinity. When compared with the value of
        $G_0 = 0.954 \pm 0.011$
                given by the GTR (32) at the standard value of
        $g_{\pi N} = 13.40 \pm 0.08$
                ,
                the interval (41) is found to be at the edge
                of consistency with the present interpretation
                of the discrepancy of the GTR (see, for
                example, Ref.
\cite{Scad}
% \bibitem{Scad} S.A.Coon and M.D.Scadron.
%       Phys. Rev. {\bf C42} (1990) 2256.
                ). This raises doubts about the discussed
                possibility to deal with the GTR discrepancy by
                lowering the value of the
        $\pi N$
                coupling constant
\cite{Pav}
                .

                By virtue of relations (39) and (40) it
                is easy to express the positions
                of the effective pole in terms of
                the cut-off parameter:
\begin{equation}
         \lambda^2 =  m^2_{\pi}
                 \left(
                        1 - g_0 \frac {\tau_m - \tau_0}
                                       {\tau_m  \tau_0}
                 \right)^{-1}
\end{equation}
                and to find the interval of
        $ \lambda $
                describing the same  region (41):
\begin{equation}
        3 m_{\pi} \leq \lambda \leq 4.20 m_{\pi}
\end{equation}
                or
\begin{equation}
        0.414 {\rm \; GeV} \leq \lambda \leq 0.580 {\rm \; GeV}
                . \nonumber
\end{equation}

                We would like to postpone the discussion how
                this meets the present experimental information
                on the position of the effective pole, in order
                to gather more details to be discussed.

                Now let us compare the FF's (21) and (23)
                in terms
                of relation (35), which requires
\begin{equation}
        \frac{G'(0)}{G(0)} = \frac{P'(0)}{P(0)}
                .
\end{equation}
                The solution of the latter equation for
        $\lambda$,
\begin{equation}
        \lambda^2 = \frac{2 \tau_m \tau_0}
                         {\tau_m  + \tau_0} \, ,
\end{equation}
                is again remarkable since it maps all the
                allowed region of
        $\tau_m$
                :
        $\tau_0 \leq \tau_m \le \infty$
                into the small enough interval of the positions
                of the effective pole:
\begin{equation}
        3 m_{\pi} \leq \lambda \leq 3 \sqrt{2} m_{\pi}
\end{equation}
                or
\begin{equation}
        0.414 {\rm \; GeV} \leq \lambda \leq 0.585 {\rm \; GeV}
                . \nonumber
\end{equation}
                This interval is practically the same as (43).
                The analogous calculation for the effective
                mass
        $\Lambda_A$
                of the dipole (24) describing the axial vector
                FF which enters the RHS of (35) provides
\begin{equation}
        \Lambda^2_A = \frac{4 \tau_m \tau_0}
                         {\tau_m  + \tau_0}
                ,
\end{equation}
\begin{equation}
        3 \sqrt{2} m_{\pi} \leq \Lambda_A \leq 6 m_{\pi}
                ,
\end{equation}
                or
\begin{equation}
        0.585 {\rm \; GeV} \leq \Lambda_A
                        \leq 0.828 {\rm \; GeV}
        . \nonumber
\end{equation}

                A straightforward comparison of this interval
                with the experimental value
\cite{Ahr}
% {Ahr} L.A.Ahrens, S.H.Aronson, P.L.Connolly et al.
%       Phys. Rev. D35 , 785 (1987).
\begin{equation}
         \Lambda_A = ( 1.032 \pm 0.036 )
                                {\rm \; GeV}
                                \, ,
\end{equation}
                results in the original impression that the
                prediction (50) is inconsistent with the
                present experimental information.To get some
                ideas on a possible interpretation
                of this contradiction, let us consider
                the properties of FF's in the asymptotic
                space-like region.

\vskip2.cm
6. LARGE MOMENTUM TRANSFERS
\vskip1.cm

                In the asymptotic region the exponential and
                the dipole FF's do not allow a comparison with
                the FF (21). When
        $\tau \gg \lambda$
                and
        $\tau \gg \tau_m$
                ,
                by equating the coefficients of the
                leading
        $ 1 / ( - \tau ) $
                terms of expressions  (21) and (23),
                one obtains
\begin{equation}
        \lambda^2 =
                \frac {  m^2_{ \pi }
                        \ln [ ( \tau_m - m^2_{\pi} ) /
                              ( \tau_0 - m^2_{\pi} ) ] }
                      { \ln \{ [  \tau_0  /
                              ( \tau_0 - m^2_{\pi} ) ]
                            [  ( \tau_m - m^2_{\pi} ) /
                                 \tau_m ] \}
                       }
                 \, .
\end{equation}

                Now the range of
        $\lambda$
                values spanned under the variation of
        $ \tau_m $
                from
        $ \tau_0 $
                to infinity is
\begin{equation}
        3 m_{\pi} \leq \lambda \leq \infty
                \, .
\end{equation}
                Contrary to the previous relations
                (46) and (42),
                where the singular character of mapping
                did not allow us to estimate
                the cut-off parameter
        $ \tau_m $
                from the value of the effective mass exceeding
                the allowed regions (43) and (47), relation
                (53) is helpful for getting an impression how
                large the values of
        $ \tau_m $
                might be.
                For the purpose of quick reference, we present
                a simple table reproducing relation
                (53) at some points:
\vskip1.cm
\begin{tabular}{|c||c|c|c|c|c|c|}
\hline
         $\lambda$ (GeV) & 0.414 & 0.450 &
                 0.500 & 0.600 &
                        0.700 & 0.800 \\
\hline
         $\sqrt{\tau_m}$ & $3 m_{\pi}$ & $3.6 m_{\pi}$ &
                 $4.5 m_{\pi}$ & $7.3 m_{\pi}$ &
                         $12 m_{\pi}$ & $20 m_{\pi}$ \\
\hline
\end{tabular}
\vskip1.cm

                The straightforward use of the common value
\cite{Nij,Thom,Cass,Dom}
% {Thom} A.W.Thomas and Karl Holinde.
%       Phys. Rev. Lett. 63 , 1025 (1989).
% {Cass} A.Cass and B.H.J. McKellar.
%       Phys. Rev. D18 , 3269 (1978).
% {Dom} C.A.Dominguez and B.J.Verwest.
%       Phys. Lett. 89B , 333 (1980).
        $ \lambda \approx 0.800 $
                GeV
                provides the value of the cut-off
        $ \tau_m $
                as large as
        $ \tau_m \approx (20 m_{\pi})^2 $
                .

                The more moderate value
        $ \lambda \approx 0.730 $
                GeV
                deduced in Ref.
\cite{Scad}
                from the axial vector FF
        $ g_A ( q^2 ) $
\cite{Ahr}
                corresponds to
        $ \tau_m \approx (13.7 m_{\pi})^2 \approx (2 m_N)^2 $
                .
                It is interesting to note that the result
\cite{Susl}
% {Susl} V.K.Suslenko, I.I.Haysak, G.I.Kolerov and
%        A.Kontantinescu.
%       In: International Seminar on Intermediate Energy
%       Physics, Moscow, 27-30 Nov., 1989, vol. 1 , pp.243-249.
                of fitting the data on the
        $ pp \rightarrow pp\pi $
                reaction at 800 MeV
\begin{equation}
        \lambda \approx 3.16 m_{\pi}  \; \Rightarrow \;
         \tau_m \approx 3.3 m_{\pi}
\end{equation}
                is precisely pointing to the region where the
                monopole approximation is being approved.

                It is premature to make definite
                statement on the value of
        $ \tau_m $
                as an experimental quantity: this should
                rather be made by independent groups possessing
                the original data.
                In this paper we would like to outline the
                ambiguities of the direct translation of
                any form-factor parameter
                to the effective mass of the
                monopole to be used in (53).

                The origin of a possible confusion might be
                clarified by examining the properties of
                the FF (21), which
                has a real chance to be
                closer to
                reality than the other
                parametrizations discussed here.
                Namely, when approximating expression (21)
                with a monopole ansatz (23), in different
                regions one inevitably obtains larger
                values  for the effective mass at larger scale
                of
        $ (-\tau) $
                than that obtained at small
        $ \tau $
                .

                The above mentioned value
        $ \lambda \approx 0.730 $
                GeV
                is obtained in Ref.
\cite{Scad}
                from the dipole parameter
        $ \Lambda_A $
                of the axial vector FF
        $ g_A ( q^2 ) $
                as
\begin{equation}
        \lambda =  \Lambda_A / \sqrt{2}
                .
\end{equation}
                In view of relation (35), this is quite
                correct for values obtained for small
        $ \tau $
                .
                However, this is not the case for the special
                example in question:
                the value (52) is obtained by fitting
                the data at the scale of
        $ \tau $
                up to
        $ \tau \approx -(1 {\rm GeV})^2 $
                .

                Certainly, the principal way is to make separate
                fittings for every choice of the FF. Since this
                has not yet been done, one might look for an
                approximate rule for quick estimations.

                To do this, let us examine the dispersion
                representation (1). Suppose that the integral
                of the imaginary part can be approximated by a
                dipole expression with the effective mass
        $ \Lambda $
                placed somewhere nearby the left end of the cut
                and that the remainder allows the approximation
                of the same kind, but with the position of its
                effective mass somewhere twice further.

                Then the fitting by a single-term dipole at
        $ ( - \Lambda^2 ) \leq \tau \leq ( - \tau_0 ) $
                should reproduce the value of the effective
                mass corresponding to the closest term and the
                value will become larger at larger scale.

                With the rough empirical rule that the
                effective mass
        $ \Lambda $
                is obtained at the scale
        $ \tau = - \Lambda^2 $,
                the prescript for a comparison of the dipole
                and monopole effective masses
\begin{equation}
        \frac { [ \Lambda^2 - m^2_{\pi} ]^2 }
              { [ \Lambda^2 - ( - \Lambda^2 ) ]^2 }
              \approx
              \frac { \lambda^2 - m^2_{\pi} }
                    { \lambda^2 - ( - \Lambda^2 ) }
\end{equation}
                should be
\begin{equation}
        \lambda \approx  \Lambda_A / \sqrt{3}
\end{equation}
                rather than (56).

                Following the conjecture of Ref.
\cite{Scad}
                that
        $ g_A ( q^2 ) $
                and the
        $\pi N\bar{N}$
                form factor should be of the same shape and
                substituting the value
        $ \Lambda_A / \sqrt{3} $
                from (52) into relation (53), one obtains
\begin{equation}
        \lambda \approx 0.596 { \rm GeV } , \; \;
        \tau_m \approx ( 7.3 m_{\pi} )^2
                .
\end{equation}
                The continuation to the small
        $ \tau $
                then results in the relative slope of
        $ g_A ( q^2 ) $
                in (35) provided by the dipole effective mass
\begin{equation}
        \Lambda^0_A \approx 0.765 { \rm GeV }   \; \;
       ( \lambda^0 \approx 0.540 { \rm GeV } ).
\end{equation}
               This is almost by 25\% smaller than (52) and
                practically consistent with the derivations of
                theoretical models, such as the Skirme model
                (see, for
                example, Ref.
\cite{Coh}
%  {Coh} T.D.Cohen. Phys. Rev. {\bf D34}, 2187 (1986).
                ).

                It goes without saying that only the direct
                fittings of the data as precise as possible at
                small
        $ \tau $
                can help to avoid the uncertainties of the
                presented speculations.

\vskip2.cm
7. CONCLUSIONS
\vskip1.cm

                In this paper we have derived a one-parameter
                expression for the
        $\pi N\bar{N}$
                form factor,
                the only input information being the existence
                of the cut in the time-like
                region, the known position of its branch point
                corresponding to the
        $ 3\pi $
                intermediate state,
                the assumption that the FF has no hard core and
                that the cut has finite length.

                The remarkable feature of the discussed FF
                in the region of small momentum transfer is
                the stability of its prediction in respect to
                the variation of the parameter (effective cut
                off):
        $ G (0) $
                is allowed to vary within
                5.5\% .
                In contrast to the result of the
                Ref.
\cite{Pav},
% {Pav} R.L.Workman, R.A.Arndt and M.M.Pavan.
%       Phys. Rev. Lett. {\bf 68}, 1653 (1992).
                where the discrepancy of the GTR was used to be
                described not by the presence of a FF, but by
                the lower
        $\pi N\bar{N}$
                coupling,
                the presented FF leaves practically no
                room for lowering the
        $g_{\pi N}$
                value (unless the value of
        $g_A (0)$
                and/or
        $F_{\pi}$
                is revised).

                The subsequent discussion about a
                possible interpretation of
                inconsistency of the predicted slope of the FF
                at
        $\tau = 0$
                with the most precise determination
\cite{Ahr}
                of
        $g_A (q^2)$,
                which, unfortunately, had been performed only
                with one fitted ansatz (dipole), displayed a
                large freedom in the continuation to the point
        $\tau = 0$
                .

                The incomplete list of problems being at
                present under study and strongly relying on the
        $\pi N\bar{N}$
                coupling constant and/or on the
        $\pi N\bar{N}$
                form factor includes:
\begin{itemize}

\item           pion-nucleon elastic and inelastic scattering;

\item           nucleon-nucleon OPE potential;

\item           few-nucleon bound states (the $S/D$ ratio of
                deuterium);

\item           pion photoproduction and electroproduction;

\item           weak
                $\pi N$
                coupling.
% PR D32 (1985) 3001.
\end{itemize}

                The FF (21), discussed in this paper,
                can in principle
                help to avoid some of the known ambiguities in
                the above mentioned problems or at least to
                outline the relevance of the
        $\pi N\bar{N}$
                coupling and/or FF to the problem in question.

\end{document}